\documentclass[aps,pra,twocolumn,superscriptaddress]{revtex4-2}

\usepackage{graphicx}
\usepackage{amssymb}
\usepackage{amsmath}
\usepackage{multirow}
\usepackage{braket}
\usepackage{mwe}
\usepackage{xcolor}
\usepackage{bbold}

\usepackage{array}
\newcolumntype{P}[1]{>{\centering\arraybackslash}p{#1}}

\begin{document}

\title{Probing the topological phase transition in the Su-Schrieffer-Heeger model using Rydberg-atom synthetic dimensions}

\author{Y. Lu}
\author{C. Wang}
\author{S. K. Kanungo}
\author{F. B. Dunning}
\author{T. C. Killian}
\affiliation{Department of Physics and Astronomy, Rice University, Houston, TX  77005-1892, USA}

\begin{abstract}

    We simulate the the Su-Schrieffer-Heeger (SSH) model using Rydberg-atom synthetic dimensions constructed, in a single atom, from a ladder of six neighboring $n\:^3S_1$ Rydberg states in which adjacent states are coupled with two-photon transitions using microwave fields. Alternating strong/weak tunneling rates, controlled by adjusting the microwave amplitudes, are varied to map out the topological phase transition as a function of the ratio of the tunneling rates. 
    For each ratio, quench dynamics experiments, in which the system is initially prepared in one of the bulk Rydberg states and then subjected to the microwave fields, are performed to measure the population evolution of the system. From the dynamics measurements, we extract the mean chiral displacement and verify that its long-time average value converges towards the system winding number. 
    The topological phase transition is also examined by probing the energy spectrum of the system in steady state and observing the disappearance of the zero-energy edge states. The results show that even a system with as few as six levels can demonstrate the essential characteristics of the SSH Hamiltonian.

\end{abstract}

\maketitle

\newpage

\section{Introduction}
\label{S:intro}
Synthetic dimensions have developed into a powerful tool for analog quantum simulations since their use was first proposed over a decade ago \cite{bcl12,opr19,hg23}. 
A synthetic dimension is constructed by encoding the spatial properties of a simulated system into the internal or external degrees of freedom of a quantum system with a high degree of experimental control. 
Such artificial systems have seen numerous realizations in the fields of photonic and atomic physics \cite{sla15,mpc15,lcd16,kbb17,mag16,vsc19,pog17,ose23}. In particular, ultracold atomic systems, which allow precise control of all degrees of freedom and provide relatively weak decoherence, have led to experimental demonstrations of synthetic dimensions in a variety of systems including hyperfine states \cite{sla15,mpc15}, atomic clock states \cite{lcd16,kbb17}, momentum states \cite{mag16,vsc19} and harmonic trap states \cite{pog17,ose23}. 
More recently, synthetic dimensions using Rydberg atomic states were also proposed \cite{opr19} and subsequently demonstrated \cite{kwl22,chc24,lwk24}, illuminating their potential, when combined with the optical tweezer arrays, in exploring interacting many-body systems \cite{opr19,fmr22,chc24}.

\begin{figure}
\includegraphics[width=8cm]{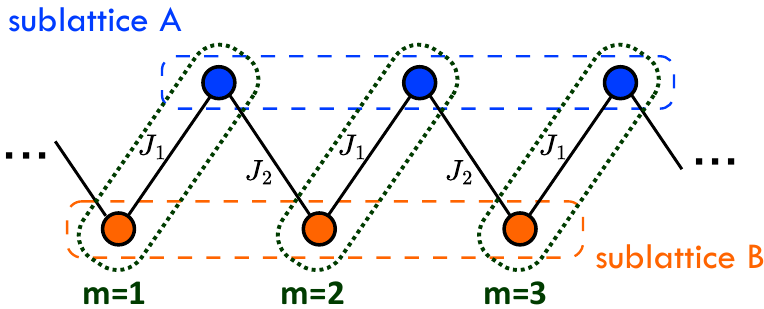}
\caption{A six-site segment of the standard 1D Su-Schrieffer-Heeger (SSH) model. The system can be divided into two sublattices as circled by the dashed lines, and tunneling only occurs between sites belonging to the different sublattices. The three unit cells making up the segment are circled by the dotted lines. The intra-cell and inter-cell tunneling rates are $\text{J}_1$ and $\text{J}_2$ respectively.
\label{fig:SSH}
}
\end{figure}

Much work with synthetic dimensions has centered on topological systems, including quantum Hall ladders \cite{sla15,cse20}, topological Anderson insulators \cite{mei18} and higher-dimensional lattices \cite{pzo18,vsc19}. Such exotic topological phases are often characterized by a global topological order, represented by a symmetry-preserving topological invariant, an example of which is provided by the Su-Schriffer-Heeger (SSH) Hamiltonian. 
The SSH Hamiltonian, inspired by studies of solitons in the polyacetylene molecule \cite{ssh79}, is a paradigmatic model in topological physics that describes a particle hopping on a one-dimensional (1D) lattice. As shown in Fig.~\ref{fig:SSH}, hopping in the standard SSH model is limited to adjacent lattice sites with two alternating tunneling rates $J_1$, $J_2$. When the system is divided into sublattices \textit{A} and \textit{B}, tunneling only exists between the two sublattices, i.e., \textit{A} to \textit{B} or \textit{B} to \textit{A}, but not within each one, i.e., \textit{A} to \textit{A} or \textit{B} to \textit{B}. Such tunneling behavior is termed ``bipartite", and the system is said to possesses chiral symmetry \cite{cts16,cds19}.

A signature behavior of the SSH model is the existence of eigen-states where population is localized on the system boundaries. This only happens when the outermost sites are weakly connected ($J_1/J_2<1$), and these boundary modes correspond to zero-energy edge states in the energy spectrum. As the tunneling ratio $J_1/J_2$ increases above $1$, these edge states disappear, and the system undergoes a topological phase transition entering the ``trivial" phase. The number of the edge states is proportional to the system winding number, which is the topological invariant generally associated with chiral systems \cite{z89,ao13}. For the 1D SSH model, the winding number takes quantized values of 1 or 0 in the topological and trivial phase, respectively. The phase transition can therefore be characterized by probing the winding number. 

The winding number can be measured from the long-time limit of a bulk observable, known as the mean chiral displacement \cite{mdc18}. This has been experimentally demonstrated using synthetic dimensions involving momentum states \cite{xgx19,mad18} and photonic states \cite{cdd17}, as well as with variants of the SSH model \cite{xgx19,mad18}.
The fact that the number of edge states is related to the topological invariant, which is a bulk property, is known as the bulk-edge correspondence.

Recently, the 1D SSH Hamiltonian was realized using Rydberg-atom synthetic dimensions consisting of six neighboring $n\:^3S_1$ states in $^{84}$Sr atoms with values of $n$ close to 60 \cite{lwk24}. Adjacent-state tunneling was realized using two-photon microwave transitions that were configured such that the system was in the topologically non-trivial phase (specifically $J_1/J_2=0.2$). Quench-type dynamics measurements were performed in which the atom was initially excited to a selected Rydberg state following which the microwave fields were turned on for a variable evolution time. Edge-to-edge long-range tunneling and bulk-population oscillations, characteristic behaviors of the topological phase, were observed in the measured population evolution \cite{lwk24}. 

In the current paper, we examine the topological phase transition that occurs as the ratio $J_1/J_2$ passes from less than to greater than one through measurements of the eigen-energies and mean chiral displacement at several different ratios of the tunneling rates in the range $J_1/J_2=0.2$\:--\:$5$. 
The mean chiral displacement, extracted from the population dynamics, reveals vanishing of the winding number as the system transitions into the trivial phase ($J_1/J_2>1$). 
The eigen-energy spectrum is probed through measurements of the Rydberg excitation spectrum in the presence of the microwave fields. Compared to earlier work \cite{kwl22}, the spectral resolution is improved significantly, and all six eigen-states are well resolved. Finite-size effects are clearly visible and even with only six states, band structure is visible. Its evolution, as the tunneling ratio passes through the critical point $\text{J}_1/\text{J}_2=1$, shows the disappearance of the zero-energy edge states, revealing the loss of topological protection.

\section{Experimental Method}
\label{s:Exp}
\begin{figure}
\includegraphics[width=8.5cm]{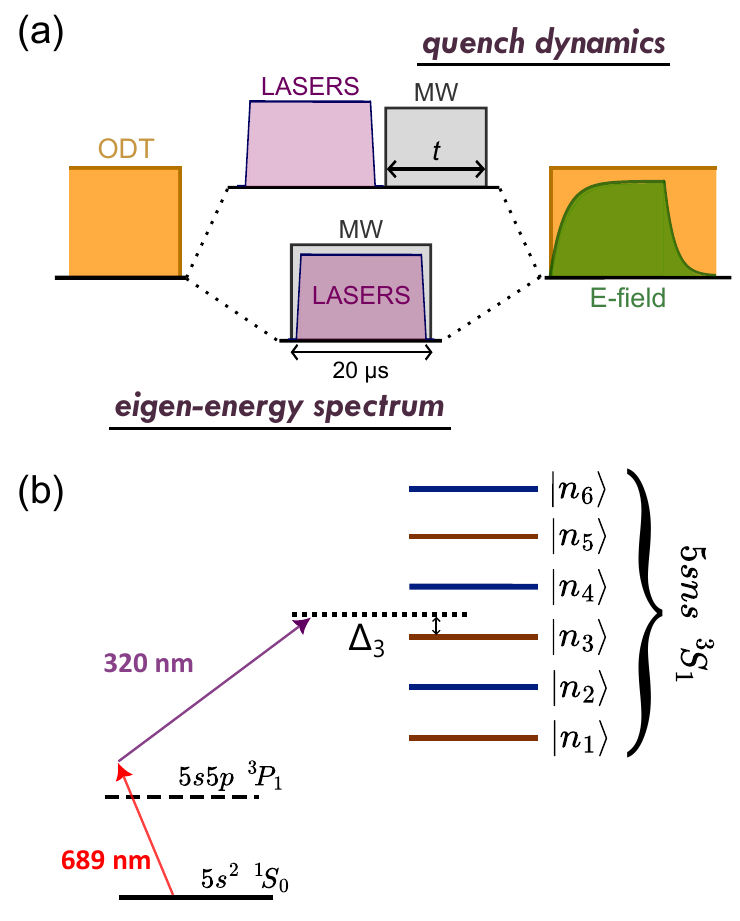}
\caption{\textbf{(a)} The pulse sequence of an experimental cycle. The optical dipole trap (ODT) is turned off during optical/microwave exposure. The ionizing E-field is ramped up after this exposure and decays to zero by the beginning of next cycle. The upper and lower branches show the exposure sequence for the quench dynamics and eigen-energy spectrum measurements respectively. \textbf{(b)} Experimental realization of the six-site SSH model. Selected $n\:^3S_1$ Rydberg states are excited using two-photon transitions through the intermediate $5s5p\:^3P_1$ state. The laser frequency can be tuned near any selected $\ket{n_i}$ state with detuning $\Delta_i$. Neighboring states are coupled by two-photon microwave transitions.
\label{fig:time_sequence}
}
\end{figure}

A cold thermal gas of $^{84}$Sr atoms is prepared through laser cooling and loaded into a crossed-sheet optical dipole trap (ODT). The trap contains $\sim\!10^5$ strontium atoms with peak density $\sim\!10^{11}$ cm$^{-3}$ and temperature $T\!=\!1$\:\textmu K (see references \cite{ssk13,emm09,dwk18} for a more detailed description of the apparatus and the cooling and trapping techniques). As shown in Fig.~\ref{fig:time_sequence}(c), the $5sns\:^3S_1$ Rydberg states are created using optical two-photon excitation via the intermediate $5s5p\:^3P_1$ state. Adjacent Rydberg levels are coupled through microwave two-photon excitation using frequencies in the range of $15-20$ GHz. A DC bias magnetic field of about $6$ Gauss is applied to split the $^3S_1$ degeneracy and selectively excite the $m=+1$ sublevel. 
The Rydberg atoms are detected by selective field ionization (SFI) in which a ramped electric field is applied in the experimental region such that different $n\:^3S_1$ states are ionized sequentially, allowing the population in each state to be detected separately. For the range of quantum numbers studied here, $n\!\sim\!60$, the ionization field reaches a maximum value of $40$\:V\:cm$^{-1}$. The liberated electrons are then directed to a microchannel plate detector where their arrival times identify the initial Rydberg state. 

An experimental cycle involves a series of steps diagrammed in Fig.\ref{fig:time_sequence}(a)(b). For measurements of the quench dynamics, the atoms are first excited to a selected $n\:^3S_1$ Rydberg level. After that, the microwave fields are turned on for a duration $t$, allowing the population to evolve under the couplings. Immediately following application of the microwave fields, the population in each Rydberg state is measured using SFI. Repeated measurements with varying microwave exposure times $t$ are used to examine the population dynamics. 

For measurements of the eigen-energy spectrum, the microwave fields are turned on before laser excitation, such that the lasers excite the atom into the microwave-dressed synthetic space instead of into the bare Rydberg states. The laser frequency is scanned near a $n\:^3S_1$ state, and the measured total excitation rate, i.e., the sum of all states present in the ionization spectrum, provides a probe of eigen-states through their overlap with the selected $n\:^3S_1$ state\cite{kwl22}. The laser linewidths, combined with the $20$\:\textmu s laser pulse widths, result in an effective laser linewidth of $\sim\!60\!-\!70\:$kHz at $n\!\sim\!60$, sufficient to easily resolve all the eigen-states of interest here.

The optical dipole trap is turned off during optical/microwave field exposure to avoid additional light shifts of the Rydberg levels. Each cold atomic sample can be used for 1000 such experimental cycles with a repetition rate of $\sim\!5$\:kHz. The average number of Rydberg atoms detected per cycle, $\sim\:0.3$, is kept low to ensure that typically less than one atom is excited to the Rydberg manifold each cycle, eliminating the effect of Rydberg-Rydberg interactions. 

Measurements were undertaken at five different tunneling ratios $J_1/J_2 = \{0.2, 0.5, 1, 2, 5 \}$ with tunneling rates of $J_1=\{160, 400, 400, 800, 800 \}$ kHz and $J_2=\{800, 800, 400, 400, 160 \}$ kHz, respectively. The tunneling rates were experimentally determined through measurement of the Autler-Townes splitting frequency for each individual transition \cite{at55,ras21}. Due to fluctuations in the microwave field strengths, the uncertainties in the measured Rabi splittings can be as large as $\sim\!20$\:kHz. The Rydberg synthetic dimension system, as depicted in Fig.~\ref{fig:time_sequence}(b), can be described with the Hamiltonian 
\begin{equation}
    \hat{H}_{\text{lattice}} = \sum^{2N-1}_{n=1}-hJ_{n,n+1}\left(\hat{c}_{n}^{\dagger}\hat{c}_{n+1} + h.c. \right) + \sum^{2N}_{n=1}\hat{c}_{n}^{\dagger}\hat{c}_{n}U_{n} 
    \label{eq1}
\end{equation}
where $2N$ is the total number of bare states and $N=3$ is the number of dimerized unit cells, and the tunneling rates alternate as $J_{n,n+1}=J_1 (J_2)$ for $n=1,3,5 (2,4)$. The on-site potential energy $U_n$ is determined by the detuning of the microwave transitions. Here we assume that $U_n=0$ for all $n$ since the microwave frequencies are adjusted to be resonant with their respective transitions. However, application of any one of the microwave fields causes AC Stark shifts that influence all other transitions.
These shifts were determined from Autler-Townes spectra measured for each individual transition within the lattice as each additional field was applied separately and were found to be as large as $\sim\!400$\:kHz, sufficient to introduce significant changes in the microwave frequencies required to resonantly couple the Stark-shifted Rydberg levels. The measured AC Stark shifts were used to adjust the applied microwave frequencies. Even with such compensation, however, the uncertainties in the AC Stark shifts can lead to detunings of up to $\sim\!50$\:kHz from resonance.

\section{Results and Discussion}
\label{s:RnD}

\subsection{Eigen-energy Spectrum}
\label{s:spectrum}
As briefly described in Sec.~\ref{s:Exp}, the eigen-states of the dressed system, i.e., Eq.~\ref{eq1}, are probed by monitoring the Rydberg excitation spectrum in the presence of the microwave fields. It can be shown that the Rydberg excitation rate using a laser tuned near resonance with the bare Rydberg states $\ket{n_i}$, before convolving with the laser linewidth, is well described by \cite{kwl22}
\begin{equation}
    \Gamma (\Delta_i) \propto \sum_{\beta}\ \lvert\braket{\beta | n_i}\rvert^2 \ \delta(\Delta_i - \epsilon_{\beta}/h)
    \label{eq_excitation}
\end{equation}
where the overlap with the selected bare state $\ket{n_i}$ is summed over all eigen-states $\ket{\beta}$ with eigen-energies $\epsilon_{\beta}$. The delta function, enforcing energy conservation, ensures that the excitation rate peaks when the laser detuning $\Delta_i$ matches one of the eigen-energies $\epsilon_\beta$ of the Hamiltonian, Eq.\ref{eq1}. 

\begin{figure}
\includegraphics[width=8.5cm]{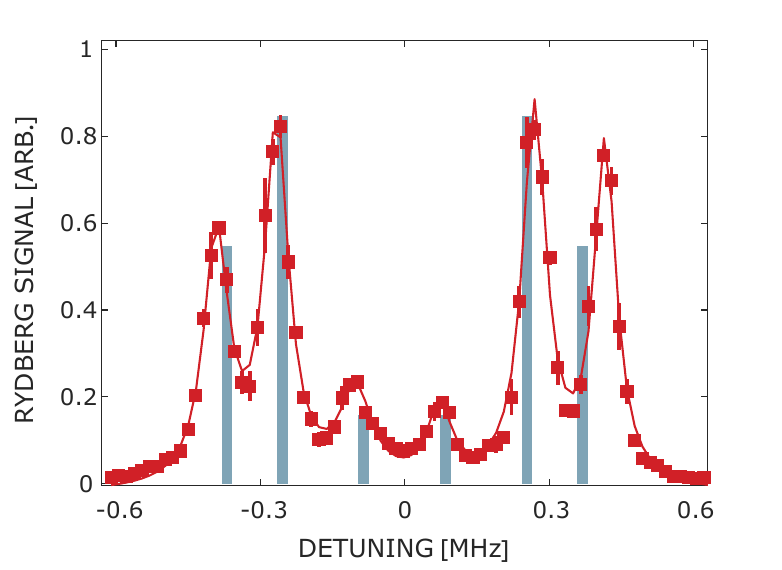}
\caption{Rydberg excitation spectrum with the laser frequency scanned across the bare state $\ket{n_2}$ for tunneling ratio $J_1/J_2\!=\!1$.
The squares are the experimental data, and the line shows the result of fitting the data using six Lorentzian functions. The vertical bars show values of $\Gamma (\Delta_2)$ calculated using Eq.~\ref{eq_excitation}.
\label{fig:spectrum_exp}
}
\end{figure}

Fig.~\ref{fig:spectrum_exp} shows a typical excitation spectrum recorded as the laser frequency is scanned across the second state $\ket{n_2}$ in the synthetic lattice using a tunneling ratio $J_1/J_2\!=\!1$. The vertical bars show the expected peak heights and positions of the eigen-states calculated by directly diagnolizing the Hamiltonian and evaluating $\Gamma (\Delta_2)$. The microwave frequencies are finely tuned, after applying the AC Stark shift compensation described in Sec.~\ref{s:Exp}, such that the general shape of the measured spectrum matches well the calculation. 
The deviations between the measured and calculated spectra can be partially attributed to uncertainties in the tunneling rates and in the microwave detunings, which are not included in the calculations.

\begin{figure}
\includegraphics[width=8cm]{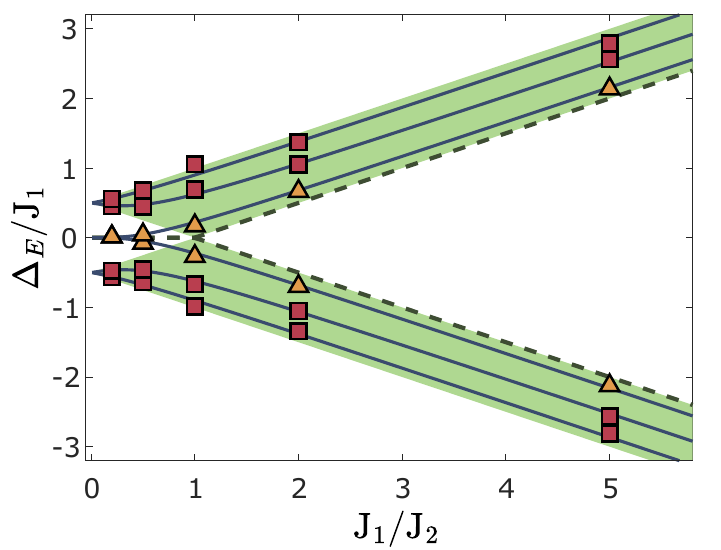}
\caption{The eigen-energies (in units of the tunneling rate $J_1$) measured from the peak positions ($\Delta_E$) in the Lorentzian fits of the Rydberg excitation spectra at different tunneling ratios $J_1/J_2$. The triangles and squares denote the edge (at small $J_1/J_2$) and bulk states, respectively, and the solid lines show the calculated results. At $J_1/J_2\!=\!0.2$, the two edge states become nearly degenerate (with $6\:$kHz energy spacing), and only one peak is resolved in the spectrum.
The shaded region indicates the calculated energy bands for a infinitely long lattice. The dashed lines show the two innermost eigen-states.
\label{fig:spectrum_ratio_plot}
}
\end{figure}

As also shown in Fig.~\ref{fig:spectrum_exp}, the measured excitation spectrum can be fit using six Lorentzian functions, each with the same width. The height of each peak gives the integrated line-strength which represents the contribution from each eigenstate.
The eigen-energies are extracted from the fit peak positions and are plotted as a function of the tunneling ratio $J_1/J_2$ in Fig.~\ref{fig:spectrum_ratio_plot}. The measured eigen-energies agree well with the calculated results despite the uncertainties in the tunneling rates and microwave frequencies. 

At small values of $J_1/J_2$ the middle two eigen-states are degenerate at zero energy, indicating the existence of topologically protected edge states. 
The phase transition that occurs as $J_1/J_2$ increases past one is clearly seen through the disappearance of the zero-energy states and the opening of a band gap.
Note that in the measured results, the critical point (where the two middle eigen-states meet at zero energy) occurs at a tunneling ratio $J_1/J_2\approx0.3$ as a result of the finite size (6 sites) of the experimental system. As can be seen in Fig.~\ref{fig:spectrum_ratio_plot}, for a infinitely long lattice, the phase transition occurs at the expected tunneling ratio $J_1/J_2\!=\!1$.

The sets of microwave frequencies and amplitudes used for each tunneling ratio $J_1/J_2$ in the spectral measurements are also used in the following studies of quench dynamics. 

\subsection{Chiral Displacement}
\label{s:chiral}
\begin{figure}
\includegraphics[width=8.5cm]{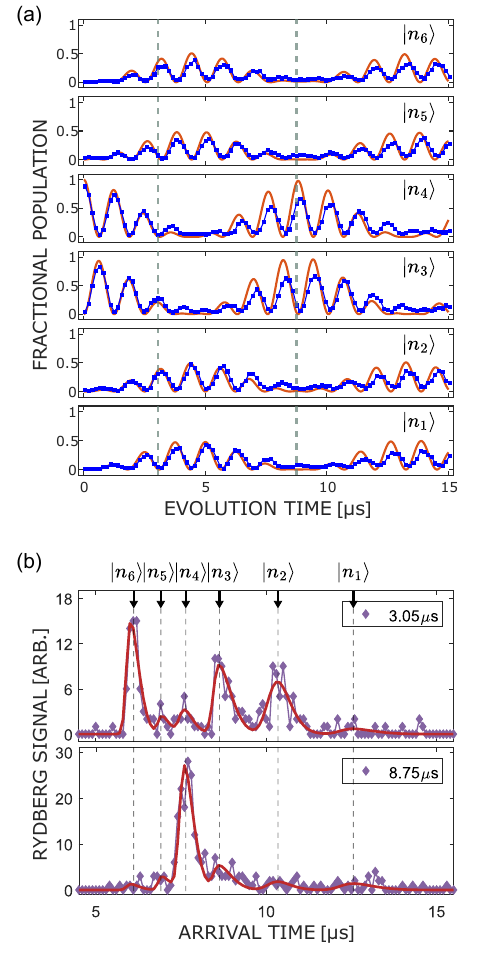}
\caption{(a) Evolution of the fractional populations in each $n\:^3S_1$ bare Rydberg state for $J_1/J_2 = 800\: \text{kHz}/ 160\: \text{kHz} = 5$ with initial excitation to the fourth state $\ket{n_4}$.  (b)(c) The SFI arrival time spectra recorded after the evolution times indicated by the vertical dashed lines in (a). The arrows indicate the Rydberg states that correspond to each feature in the arrival time spectrum.
\label{fig:dynamics_data}
}
\end{figure}

The results of a typical quench dynamics measurement are shown in Fig.~\ref{fig:dynamics_data}(a) for a tunnling ratio of $J_1/J_2 = 5$. As the microwave couplings are turned on, the population that is initially localized on the $\ket{n_4}$ site quickly diffuses into all other sites. The significant amount of population that tunnels into the two outermost sites demonstrates that the system is in the topologically trivial phase for which there are no protected edge states. This is to be expected given that the SSH Hamiltonian only transitions to the topologically non-trivial phase when $J_1/J_2 < 1$.

Assessment of which topological phase a chiral system is in, however, does not necessarily involve direct observation of topological behaviors such as the presence of protected edge states.
For a 1D chiral system such as that described by the SSH Hamiltonian, the chiral operator $\hat{\Gamma}$ can be defined in the block-diagonal form 
\begin{equation}
    \hat{\Gamma} = \begin{pmatrix}
                        \mathbb{1} & 0\\
                        0 & -\mathbb{1}
                        \end{pmatrix}
\end{equation}
where the two identity matrices run through the sublattices A and B, respectively. One can also define the position operator $\hat{m}$ with $m$ labeling the unit cells shown in Fig.~\ref{fig:SSH}. It has been shown \cite{mdc18} that the expectation value of the composite `chiral position' operator $\braket{\hat{\Gamma m}}$ reveals the system winding number $\mathcal{W}$ via the bulk dynamics. Specifically, with the initial population localized on a single site in the standard SSH model, we have $\braket{\hat{\Gamma m}(t)} = \mathcal{W}/2 + ...$, where $...$ denotes oscillating terms that are averaged out in the long-time limit. For convenience, here we define the mean chiral displacement value as $C(t) = 2\braket{\hat{\Gamma}\hat{m}(t)}$ such that $C(t)\cong\mathcal{W}$ at late times, and it takes the form, for our 6-state SSH system,
\begin{equation}
    C(t) = 2\braket{\hat{\Gamma}\hat{m}(t)} = 2\:[ P_1 - P_2 + 2\:P_3 - 2\:P_4 + 3\:P_5 - 3\:P_6]
    \label{eq2}
\end{equation}
where $P_i$ is the probability of being in the $i$-th bare Rydberg state. This can be directly evaluated from the measured population dynamics.

\begin{figure}
\includegraphics[width=8cm]{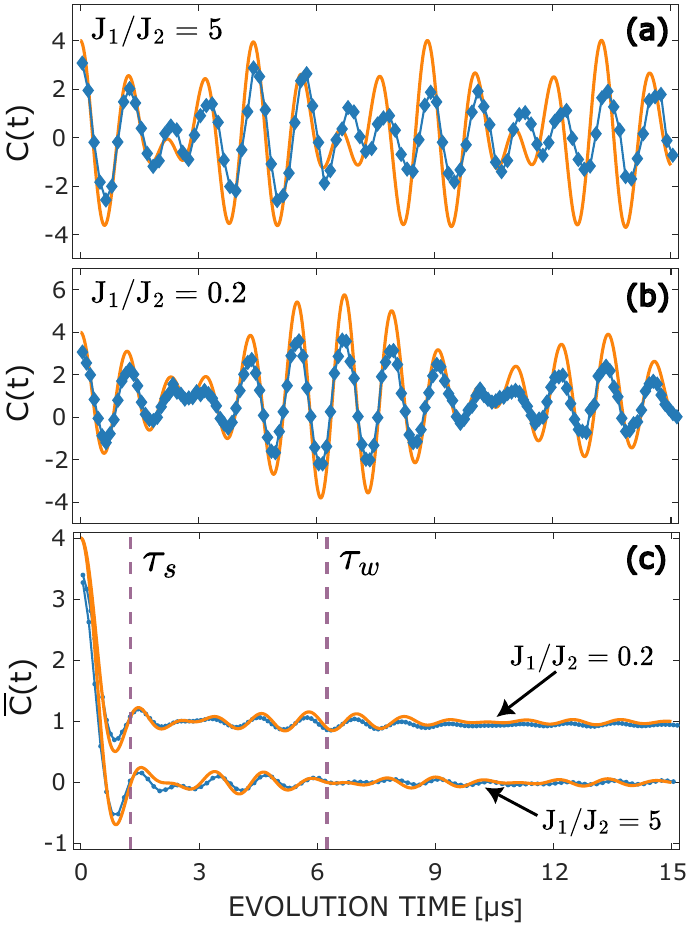}
\caption{(a) Mean chiral displacement $C(t)$ as a function of the evolution time $t$ for $J_1/J_2 =800\text{\:kHz}/160\text{\:kHz}= 5$. The blue connected dots are the experimental data. The orange solid curves are the calculated results. (b) for $J_1/J_2 = 160\text{\:kHz}/800\text{\:kHz} = 0.2$. (c) The cumulative average $\overline{C}(t)$ for the two tunneling ratios. Vertical dashed lines mark the characteristic periods $\tau_w$, $\tau_s$ corresponding to the tunneling rates (see text.)
\label{fig:chiral_raw_2}
}
\end{figure}

Figs.~\ref{fig:chiral_raw_2}(a)(b) show the mean chiral displacement $C(t)$ evaluated from the dynamics data for $J_1/J_2=5$ and $J_1/J_2=0.2$ respectively. The measured values of $C(t)$ agree reasonably well with theoretical predictions obtained from direct diagnolization of the SSH Hamiltonian (Eq.\ref{eq1}). The dynamics of $C(t)$ show oscillatory behavior as expected. In order to average out the oscillating terms in $C(t)$, the cumulative time average, i.e., the average over all past time $\overline{C}(t)=1/t\cdot\int_0^{t'} C(t')\:dt'$, is taken and shown in Fig.~\ref{fig:chiral_raw_2}(c).

The oscillations present in $C(t)$, and in the original population dynamics, have characteristic periods naturally determined by the tunneling rates $J_1$, $J_2$, each of which has an associated time scale $\tau_w = 1/J_w$ and $\tau_s = 1/J_s$, with $J_w$, $J_s$ being the weaker and stronger of $J_1$, $J_2$, respectively. The measurements of $C(t)$ for the two different tunneling ratios $J_1/J_2=0.2 \text{ and } 5$ share exactly opposite strong/weak Rabi frequencies. As indicated by the vertical dashed lines in Fig.~\ref{fig:chiral_raw_2}(c), the initial transient, which is determined by the initially localized state, is quickly averaged out within $t\simeq\tau_s$. Beyond the weak-tunneling time scale $t\gtrsim\tau_w$, most of the oscillations in $\overline{C}(t)$ have died out, and $\overline{C}(t)$ converges to 0 or 1, which is consistent with the expected values of winding number for the trivial ($J_1/J_2\!=\!5$) and topological ($J_1/J_2\!=\!0.2$) phases, respectively.

Measurements of $C(t)$ and the time average $\overline{C}(t)$ are repeated for five different tunneling ratios $J_1/J_2$ with the same range of evolution time $t\!=\:0-15\!$ \textmu s. At each ratio, we take the final ($t=15\!$ \textmu s) value of $\overline{C}(t)$ as the long-time average of $C(t)$, i.e., we approximate $\overline{C}(t\rightarrow\infty) \approx \overline{C}(t\!=\!15$\textmu s$)$. The $15\!$ \textmu s averaging time is longer than the characteristic times $\tau_w, \tau_s$ for any of the tunneling rates used here. 

\begin{figure}
\includegraphics[width=8.5cm]{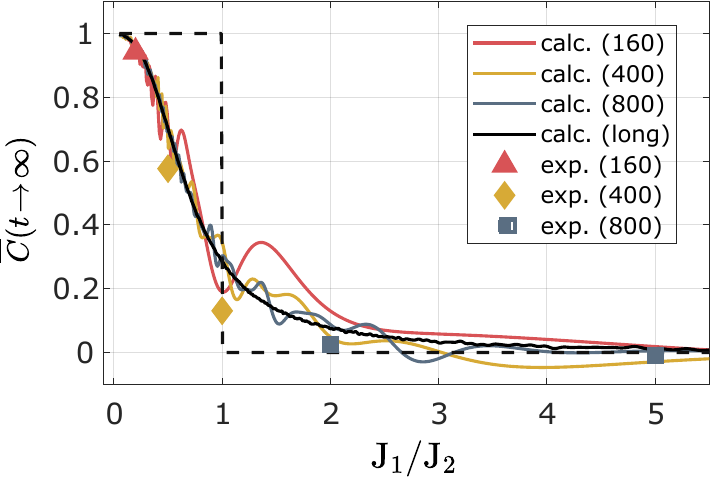}
\caption{The long-time average of the mean chiral displacement $\overline{C}(t\rightarrow\infty) \approx \overline{C}(t=15$\textmu s$)$ as a function of the tunneling ratio $J_1/J_2$. The symbols show the measured results obtained using the values of $J_1$ indicated in parentheses. The solid curves show the corresponding calculated values. The black solid line shows the limiting case of a very long-time average, and the dashed line is the behavior expected for an infinitely long lattice.
\label{fig:chiral_ratio_plot}
}
\end{figure}

The measured values of $\overline{C}(t)$ are plotted in Fig.~\ref{fig:chiral_ratio_plot} as a function of the tunneling ratio. The long-time limit of the chiral displacement $C(t)$, which provides a measure of the system winding number $\mathcal{W}$, exhibits a clear transition from 0 to 1 as the tunneling ratio $J_1/J_2$ decreases through one, further demonstrating the trivial-to-topological phase transition seen in the eigen-energy spectrum. As a result of the finite size ($6$ sites) of our experimental system, the data points exhibit a smooth crossover near the critical point $J_1/J_2=1$ rather than the step function expected in the case of an infinitely long chain (see Fig.\ref{fig:chiral_ratio_plot}).

The data points and calculations are further categorized in Fig.\ref{fig:chiral_ratio_plot} into groups with different $J_1$ (160 kHz, 400 kHz, 800 kHz). In particular, the theoretical calculations at the three values of $J_1$ share a common general trend with which the measured data points agree reasonably well. The value of $\overline{C}(t)$, as obtained from time averaging $C(t)$, generally depends on both the characteristic periods $\tau_w, \tau_s$ (equivalently $J_w, J_s$) and the averaging time. Since the latter is fixed at $15$\:\textmu s, changes in $\tau_w$, $\tau_s$ lead to variations in the number of periodic oscillations that are being averaged over. This is reflected in the smaller oscillation amplitudes in the calculated curves in Fig.\ref{fig:chiral_ratio_plot} as the tunneling rates increase.
At $J_1\!=\!800\:$kHz, the calculated curve is already relatively smooth with only small oscillations as the long-time limit is approached, demonstrating that cumulative averaging efficiently extracts the winding number from the chiral displacement measurements.

\section{Conclusions}
\label{s:Conclusions}
The topological phase transition present in the SSH Hamiltonian has been explored by measuring the eigen-energy spectrum and, through the mean chiral displacement, the winding number using a 6-state finite 1D SSH model based on Rydberg atom synthetic dimensions. The measured results agree well with theoretical predictions and capture the essential characteristics of the system near its phase transition. In particular, it is shown that long-time averaging of the mean chiral displacement can provide a reliable measure of the winding number even for limited system size and sampling time, both of which can be limited in experimental settings due to technical challenges and decoherence. 

The present experimental scheme can be extended to larger system using $8$-$10$ Rydberg states or more. It is also possible, by directly coupling non-adjacent states, to form closed loops and plaquettes, which can create complex lattice geometries and realize artificial gauge fields \cite{chc24,cmr14,cbm19}. Rydberg synthetic dimensions can also be combined with arrays of single Rydberg atoms in closely spaced optical tweezers \cite{bl20} to exploit Rydberg-Rydberg interactions and realize multi-dimensional systems such as quantum strings and membranes \cite{fmr22,stw19}.

\begin{acknowledgments}
The authors are pleased to acknowledge valuable discussions with K. R. A. Hazzard during the course of this work which was supported by the NSF under Grants No. 1904294 and No. 2110596.
\end{acknowledgments}

%\bibliographystyle{apsrev4-2}
%\bibliography{Bibliography.bib}


\end{document}